\documentclass[a4paper]{article}
\usepackage{graphicx}
\usepackage{fullpage}
\usepackage{latexsym}
\usepackage{mathrsfs}
\usepackage{amssymb,amsbsy}
\usepackage{amsfonts}
\usepackage{caption}
\usepackage{subcaption}
\usepackage{float}
\usepackage{wrapfig}
\usepackage{epstopdf}
\usepackage{multirow}

\usepackage{amsmath}
\usepackage{authblk}
\usepackage{natbib}
\def\barr{\begin{array}}
\def\earr{\end{array}}
\def\berr{\begin{eqnarray}}
\def\err{\end{eqnarray}}
\def\berrno{\begin{eqnarray*}}
\def\errno{\end{eqnarray*}}
\def\be{\begin{equation}}
\def\ee{\end{equation}}

\def\barr{\begin{array}}
\def\earr{\end{array}}
\def\berr{\begin{eqnarray}}
\def\err{\end{eqnarray}}
\def\berrno{\begin{eqnarray*}}
\def\errno{\end{eqnarray*}}
\def\be{\begin{equation}}
\def\ee{\end{equation}}

\usepackage{comment}

\date{}
\begin{document}

\title{Testing a Prototype 1U CubeSat on a Stratospheric Balloon Flight}

\author[1,2]{S. Akaash\footnote{E-mail:s.akaash2015@vitalum.ac.in\\At time of the work, S. Akaash was a project student at the Indian Institute of Astrophysics, Bangalore.}}
\author[2]{Bharat Chandra} 
\author[2]{Binukumar G. Nair} 
\author[2]{Nirmal K.} 
\author[2]{Margarita Safonova} 
\author[2]{Shanti Prabha} 
\author[2]{Rekhesh Mohan}   
\author[2]{Jayant Murthy}  
\author[1]{Rajini G.K.} 

\affil[1]{Vellore Institute of Technology, Vellore}
\affil[2]{Indian Institute of Astrophysics, Bangalore 560034} 

\maketitle

\begin{abstract} 
High-altitude balloon experiments are becoming very popular among universities and research institutes as they can be used for testing instruments eventually intended for space, and for simple astronomical observations of Solar System objects like the Moon, comets, and asteroids, difficult to observe from the ground due to atmosphere. Further, they are one of the best platforms for atmospheric studies. In this experiment, we build a simple 1U CubeSat and, by flying it on a high-altitude balloon to an altitude of about 30 km, where the total payload weighted 4.9 kg and examine how some parameters, such as magnetic field, humidity, temperature or pressure, vary as a function of altitude. We also calibrate the magnetometer to remove the hard iron and soft iron errors. Such experiments and studies through a stratospheric balloon flights can also be used to study the performance of easily available commercial sensors in extreme conditions as well. We present the results of the first flight, which helped us study the functionality of the various sensors and electronics at low temperatures reaching about $\pm -40^{\circ}$. Further the motion of the payload has been tracked throughout this flight. This experiment took place on 8 March 2020 from the CREST campus of the Indian Institute of Astrophysics, Bangalore. Using the results from this flight, we identify and rectify the errors to obtain better results from the subsequent flights.
\end{abstract}

{\bf Keywords}: high-altitude balloons, attitude, stratosphere, CubeSat, payload motion, magnetometer. 

\section{Introduction}

High-altitude balloon experiments are becoming very popular as they can be used for scientific research at fractions of the cost of space projects. Because of the low cost, they are perfect for both academic and educational institutions. Besides that, at high altitudes we escape most of the atmosphere, thus avoiding the need to resort to adaptive optics as in the case of ground-based telescopes \citep{Ref-1}. Since balloons are launched to the altitudes of about 25--50 kilometers, the instruments are exposed to a harsh environment, where the temperature reaches about $-80^{{\circ}}$C and pressure about 1 millibar, thus balloons serve as a good platform to test the functionality of the components before launching into space. Space payloads for UV imaging and observations can be tested for their performance on high-altitude balloons \citep{{Ref3},{Ref2},{Ref5}}. Pointing systems are necessary for the study of astronomical sources, and such systems can also be studied and tested using balloons. The design of the balloon-borne pointing system, which can be used to point and track objects with an accuracy of $\pm 0.13^{\circ}$ and $\pm 0.28^{\circ}$, respectively, is described in \citep{Nirmal}.

Among the various instruments that are being used for space research, CubeSats are becoming very popular among academic and research institutions. Their beginning was marked by a collaboration between California Polytechnique State University and Stanford University in 1999. The purpose of such miniature satellites was to make research in space be accessible to the universities. Nowadays, even high schools have started sending CubeSats to space. These satellites are mainly used for climate monitoring and changes, biological science, the study of near-Earth objects, planetary science, space-based astronomy, stellar imaging and heliophysics \citep{C0}. 

 Like other instruments discussed above, CubeSats also need to be tested before they are deployed into space. A study shows that though the success rate of the CubeSats slowly shows an increasing trend, it still has a failure rate of about 22\%. This can be improved considering that more and more of them are being launched over the years \citep{C1}. Thus, it might be better to test them and analyze their performance on a high-altitude balloon before actually launching them in space. 
 Testing CubeSats can reveal some of the problems in the subsystems. For example, the research article by Norwegian University of Science and Technology (NTNU) describes such an experiment testing the functioning of UHV and VHF radio on a meteoballoon, which revealed some of the thermal issues and problems with utilizing HAM radios in a CubeSat \citep{C3}. In \citet{C3test}, it was shown that there was a loss of contact with the balloon at very high altitudes. The study suggested that high-speed winds, with a speed of about 250 miles per hour, caused this causing the swinging of the antenna. A study from another flight describes the overall preliminary performance test of the communication system on the student CubeSat NAVIS \citep{C3_5}. Through the data obtained from a balloon flight, it was inferred that a 1U CubeSat is enough to receive signals from the Automatic Identification System around Greenland. Testing CubeSats on balloons can also prove to be a cheap and an excellent opportunity for students to get hands-on experience for building space instruments which are to be launched in space \citep{C4}.

In 2011, the High-Altitude Ballooning (HAB) program was initiated at the Indian Institute of Astrophysics (IIA) with the primary purpose of developing and flying low-cost scientific payloads on balloon-borne platforms and, also develop instruments that can operate on a range of near-space platforms, including CubeSats, minisatellites, or even in space missions. The results of initial tethered flights, performed at the IIA are described in \citet{Ref1}. \citet{Ref0} provides a detailed explanation of stratospheric balloon experiments, and describes the results from 9 such scientific experiments performed by the IIA balloon research group. Such experiments can be used to study the planets in our Solar System, the Moon, comets, and even atmospheric emission lines \citep{Ref2}.

In this work, we designed a prototype 1U CubeSat to verify the functionality of various sensors which are used to measure such parameters as pressure, temperature, magnetic field, humidity. We describe here the ground-based tests, tethered flight tests, and finally, the results of the stratospheric balloon flight performed on March 8, 2020.

\section{CubeSat Design}

\begin{figure}[ht!]
\centering
\includegraphics[scale=0.7]{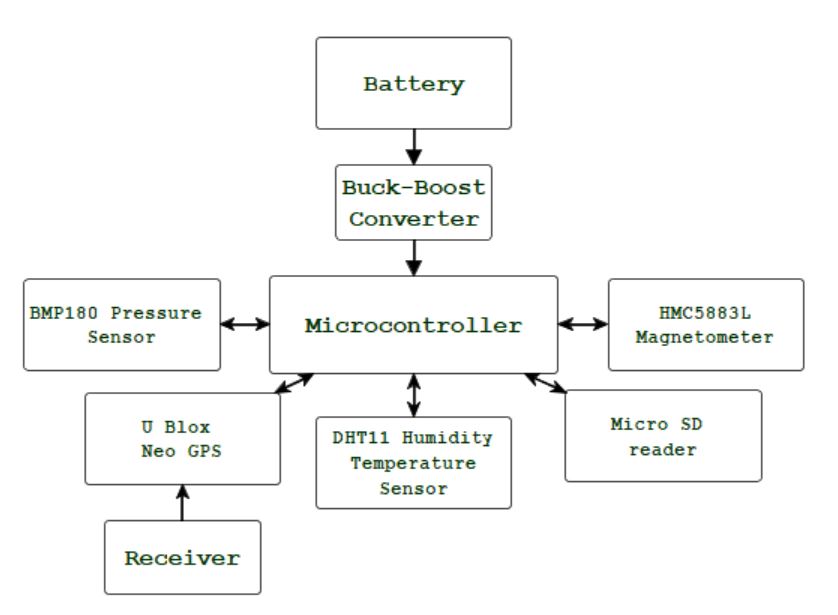}
\caption{A simple flowchart of the CubeSat.}
\label{flowchart}
\end{figure}

The payload is a 1U CubeSat which has the dimensions of about $10\times10\times10$ cm. The aluminium frame used was purchased from \textit{Interorbital Systems}\footnote{https://www.interorbital.com/}. For this experiment, the CubeSat was designed such that the frame contains three layers of PCB  connected with each other for proper functioning. The first layer is the microcontroller section, which acts like the brain of the CubeSat. The microcontroller is used to give commands from the various sensors and receive the information from them. The second layer is the experimental layer, which contains the various sensors and electronic modules essential for the experiment. The third layer is the power supply layer, which is used to supply power to the CubeSat. A flowchart indicating the connectivity of the various components has been shown in Figure~\ref{flowchart}.

\subsection{Microcontroller and Power Section}

\begin{figure}
    \centering
    \includegraphics[scale=0.6]{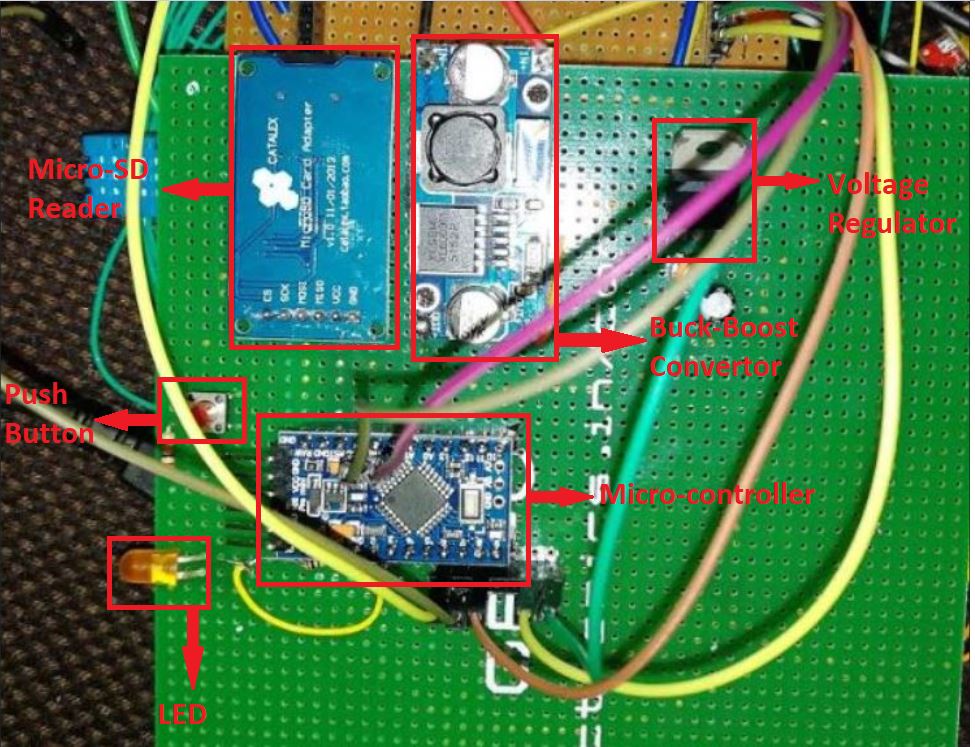}
     \caption{Microcontroller section of the  CubeSat built using perforated boards, with various components indicated.}

    \label{micro}
\end{figure}

Perforated boards were used to build the prototype. The microcontroller we have used in the design is ATmega328 with Arduino firmware. The microcontroller sends commands and acquires data from the various sensors on the CubeSat. There are 14 input-output pins, and the microcontroller board produces a 5V output which is the voltage required by most sensors. However, we also have a buck-boost converter so that we can supply power to devices that run at other voltages. This buck-boost converter is connected to the battery which converts the unregulated voltage to a regulated 5V supply. This steady supply is given to the microcontroller. We also have a micro SD card module which stores the recorded values from all the sensors in the experimental section. Sometimes it so happens that when the SD cards are suddenly removed without turning off the power supply in the CubeSat, the data gets corrupted. To ensure this doesn't happen, a safety power-off, a push button has been placed to stop the writing of data to the SD card and to make sure that it can be removed safely. The red LED is used to indicate if the data is being written to the SD card or not. Voltage regulators and filters were used for ensuring that sudden changes in voltages do not damage the working components. This entire Microcontroller subsection is represented in Fig.~\ref{micro}. Finally, the buck-boost converter is powered by a 3.7 V 5200 mAh battery. Separate insulation was not provided and whatever insulation was provided by the manufactures was used. This battery was placed inside the CubeSat as the third layer.

\subsection{Experimental Section}

We describe here the sensors and electronic components that measure parameters of interest, such as pressure, magnetic field, temperature, humidity and location. It is especially very important to calibrate the magnetometer before using it. 
 
 \subsubsection{Magnetometer Calibration}
 
\begin{figure}[ht!]
\centering
 \includegraphics[width=.45\textwidth]{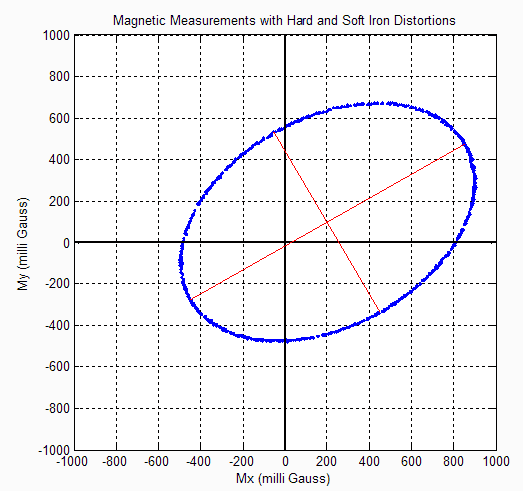}
 \hskip 0.2in
\includegraphics[width=.45\textwidth]{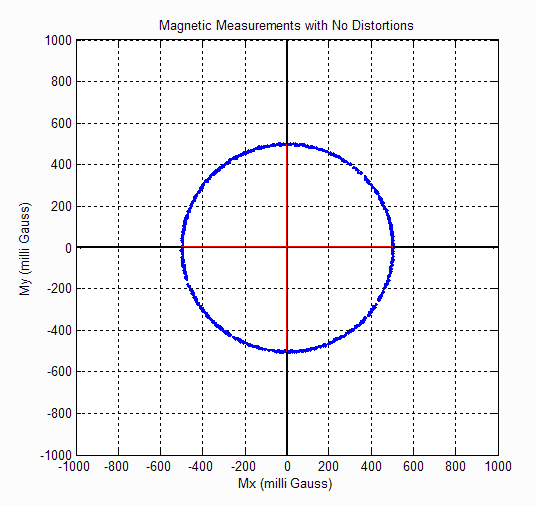} 
\caption{The left image shows the magnetic field obtained when there is a distortion due to hard and soft iron errors. The right image shows the indicates calibrated magnetic field which is free from hard and soft iron errors. (Taken from https://www.vectornav.com/support/library/magnetometer)}
\label{errors}
\end{figure}

The magnetometer used here was the HMC-5883L triple-axis magnetometer. Change in resistance of the Ni-Fe material in the magnetometer is detected through the bridge circuit which is then used to estimate the value of the magnetic field. The resistance of the material changes because the Earth’s magnetic field affects the flow of charges in the conductor. This sensor works based on the I2C communication protocol and the $X$, $Y$ and $Z$ values of the obtained magnetic field has to be obtained from the data stored in the 8-bit registers. However the obtained value is not the true magnetic field as there are instrumentation errors and hard and soft iron errors which have to be corrected. Instrumentation errors can be caused due to some errors in fabrication of the device which can introduce scale factor errors, which are errors caused in the proportionality constants relating the input and output or offsets, which introduces some bias in the magnetic field. On the other hand, hard iron effect introduces a bias in the reading due to the presence of permanent magnetic materials, and soft iron effects are caused by external magnetic fields \citep{I1}. An example of the effects of these errors are shown in Fig.~\ref{errors}.
 
\begin{figure}[ht!]
\centering
 \includegraphics[height=0.4\textwidth,width=.35\textwidth]{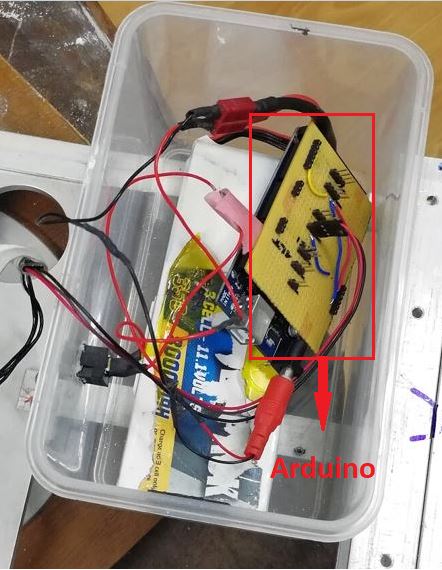}
 \hskip 0.2in
\includegraphics[height=0.4\textwidth,width=.35\textwidth]{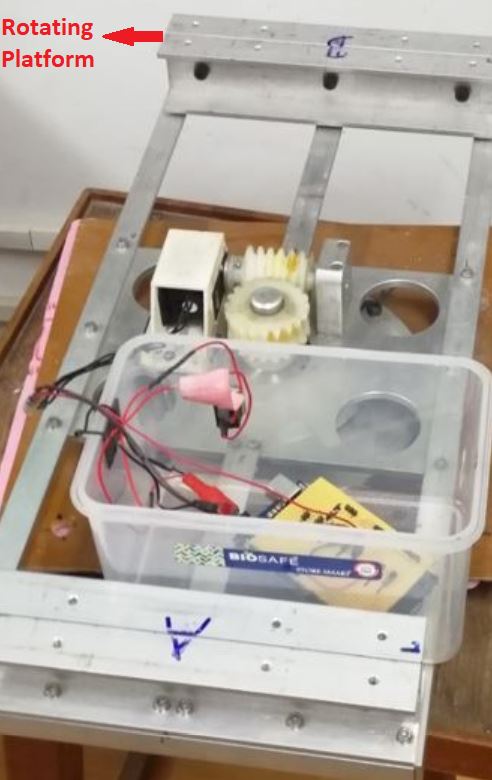} 
\caption{ {Images show the setup which can be used to measure the repeatability and linearity. The platform rotates with the help of a motor which is controlled with the help of Arduino. The magnetometer is fixed on this system which remotely sends the data to a Raspberry Pi. 
}}
\label{setup1}
\end{figure}

\begin{figure}[ht!]
\centering
 \includegraphics[width=.45\textwidth]{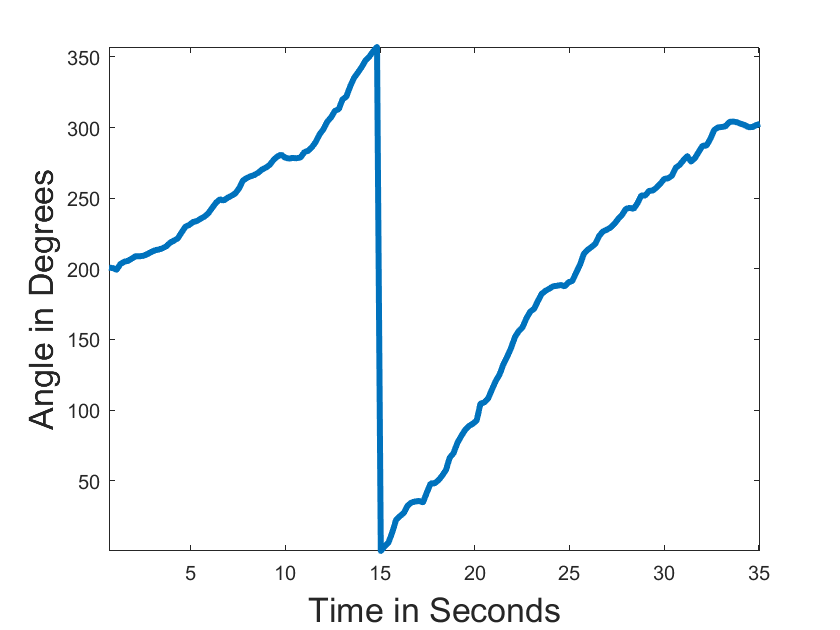}
 \hskip 0.2in
\includegraphics[width=.45\textwidth]{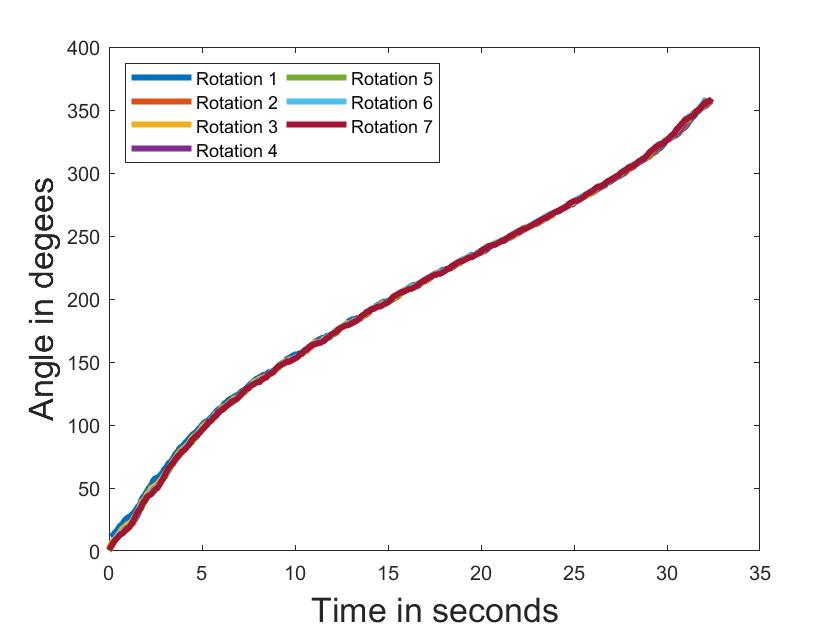} 
\caption{Left plot shows the linearity test performed by a uniformly rotating the magnetometer by hand. Though it is not exact, it does give something close to a linear result. Right plot displays the results of the repeatability test, showing that the magnetometer output is quite repeatable and that it is functioning properly.}
\label{setup}
\end{figure}

Next, we perform the linearity test. This means that, as we change the input, we obtain a linearly proportional output. In this case, the test was performed by slowly changing the azimuth by rotating the magnetometer with time. The azimuth is given by: \\
\begin{equation}
    \phi=\arctan(B_{y},B_{x})\,.
\end{equation}
To perform the linearity test, we rotated the magnetometer by hand at uniform speed. The linearity test can also be used to check the proper functioning of the magnetometer. Sometimes, it might so happen that the output fluctuates very badly indicating that the magnetometer is faulty. In fact, the first magnetometer that was tested did not pass this test, meaning that it was faulty.
To perform the repeatability test, we need to rotate the magnetometer about various axes. Both the linearity and repetability tests are shown in Fig.~\ref{setup}. So we fix it on a pointing system \citep{Nirmal} developed at the Indian Institute of Astrophysics, and rotate it with the help of a DC motor which is in turn controlled by an independent Arduino UNO microcontroller board Fig.~\ref{setup1}. The data obtained from the magnetometer was sent remotely to a computer. To do this, we had used a Raspberry Pi 3A+ to establish a remote connection with the magnetometer. We can also verify the linearity through the repeatability test by looking at the input (time) vs output (azimuth) response.

\begin{figure}[ht!]
\centering
 \includegraphics[height=0.4\textwidth, width=0.45\textwidth]{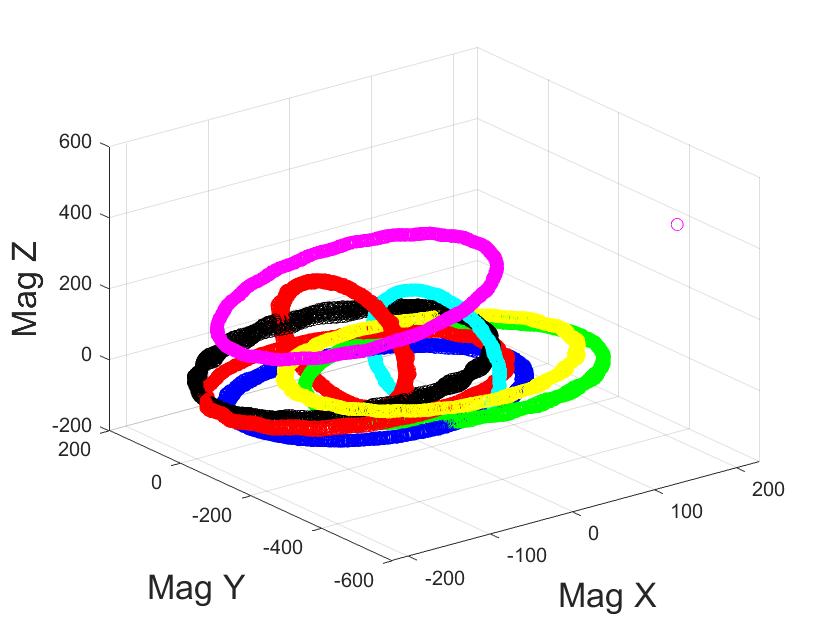}
 \hskip 0.2in
\includegraphics[height=0.4\textwidth, width=0.45\textwidth]{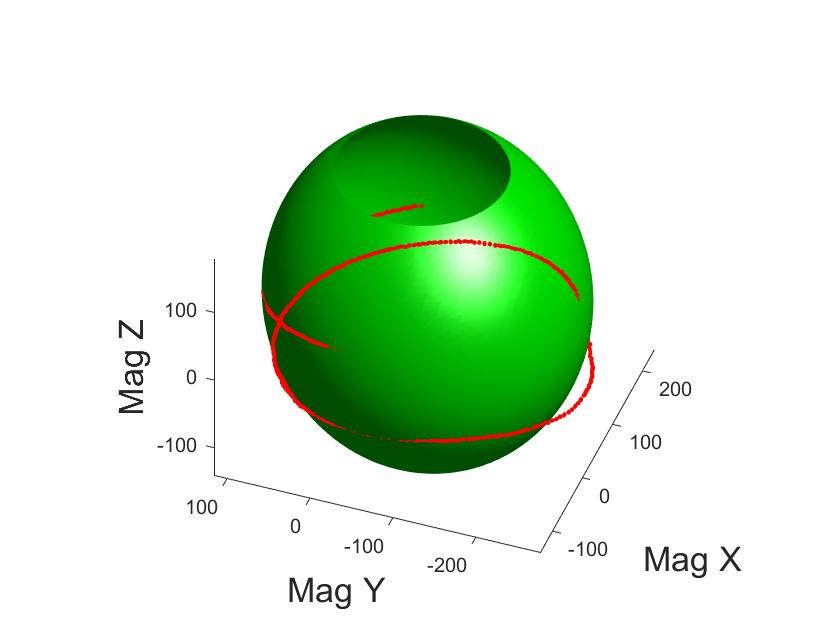} 
\caption{{\it Left}: This image shows the uncalibrated magnetic field. It was obtained by keeping the magnetometer at various possible orientations. The different colours in the graph represent the rotations performed with different orientations of the magnetic field. We can see that there are several offsets in it.Mag X, Mag Y and Mag Z axes represent the total magnetic field with respect to the directions mentioned on the HMC5883L magnetometer and is measured in micro Tesla. {\it Right}: This image shows the results of the calibration which required using the ellipsoid-fitting algorithm in MATLAB. }
\label{ellipsoidfitting}
\end{figure}

We tried several methods to calibrate the magnetometer data. However, none of them gave a properly calibrated output. After some trial and error, we found that the ellipsoid-fitting algorithm\footnote{https://in.mathworks.com/matlabcentral/fileexchange/23377-ellipsoid-fitting?requestedDomain=www.mathworks.com} was the best to calibrate the magnetometer\footnote{ Detailed explanation: https://teslabs.com/articles/magnetometer-calibration/}. The output after calibration is shown in Fig.~\ref{ellipsoidfitting}. After some rigorous calculations, we arrive at the following equation: 

\begin{equation}
 h_m=Ah+b\,,
\label{finaleqnmagfield}
\end{equation}
where $h_m$ denotes the obtained magnetic field and $h$ denotes the true magnetic field. The constants $A$ and $b$ are determined by the ellipsoid-fitting algorithm, which was performed in MATLAB. To determine the constants, we need to obtain some raw data from the magnetometer by ensuring that the axis is pointed at various angles. The constants $A$ and $b$ were found to be:
\begin{center}
$A=
\begin{pmatrix}
1.290&0.1457&0.0535\\
-0.1647&1.7430&0.03602\\
-0.4739&0.4198&1.7794\\
\end{pmatrix}$ \,,\quad $ b=
\begin{pmatrix}
-0.345\\
-20.563\\
-0.418\\
\end{pmatrix}$
\,.
\end{center}

\subsubsection{Other sensors}

This section also contains the Ublox Neo-6M GPS\footnote{https://www.u-blox.com/en} which needs to be supplied with a constant voltage of 3.3V. The GPS module needs to lock on to at least 4 satellites to get the latitude and longitude, and an additional satellite to get the altitude. GPS data contains 3 different types of information. A pseudo-random code tells us to which satellites the GPS is locked on. Ephemeris data tells the position of that particular satellite, its health, date, and time. Almanac data shows orbital data of every satellite. Using the GPS data, we find out the latitude, longitude, and altitude. However, there are a lot of other data like flight velocity and satellite details which can be obtained. Care should be taken to ensure that the baud rate of the GPS is perfectly set so that communication with the microcontroller is not affected. The microcontroller must be programmed such that the data is requested from the GPS during the wait time of the GPS. If the data is requested during the processing time -- the time when the GPS is acquiring the data from the satellites, it might lead to invalid or corrupted data. 

The DHT-11 temperature-humidity sensor was used for measuring both temperature and humidity. The humidity is measured by the level of water vapour in the atmosphere. There are two electrodes inside which form a capacitive system. When the water molecules stick to the surface, ions are liberated thus changing the conductivity, and this can be used to measure the water level and thus humidity. Temperature is obtained from the temperature-sensitive voltage and current characteristics of a diode. When two identical transistors are operated at a constant ratio of collector current densities, the difference in base-emitter voltages is directly proportional to the absolute temperature. The temperature sensor in the DHT-11 works only in the small range:  $0^{\circ}$C to $ 50^{\circ}$C while the humidity sensor works between  $10 \%$ or $20 \% $ to $ 90 \%$.  

The pressure sensor used was BMP 180 pressure sensor and uses the I$^{2}$C communication. A pressure sensor uses a piezo-resistive sensor to measure the pressure. This piezo-resistive sensor is connected in a wheat stone bridge. Now when pressure is applied to it, its resistance changes and because of this, the output in the bridge changes, thus indicating the pressure. BMP180 is also useful as it has a very good temperature sensor which works from $-40^{\circ}$C to $ 80^{\circ}$C. Using the pressure reading, we can also estimate the altitude from Eqn. (\ref{pressure})\footnote{https://www.circuitbasics.com/set-bmp180-barometric-pressure-sensor-arduino/}:

\begin{equation}
A=44330 \left(1-\frac{p}{p_{o}}\right) ^ {\frac{1}{5.255}}\,,
\label{pressure}
\end{equation}
where $A$ is the estimated altitude above the launch point, $p$ is the pressure obtained and $p_{o}$ is the pressure at the point of launch. 

\section{Description of the experiment}

\begin{figure}[ht!]
\centering
 \includegraphics[width=.40\textwidth, height=0.45\textwidth]{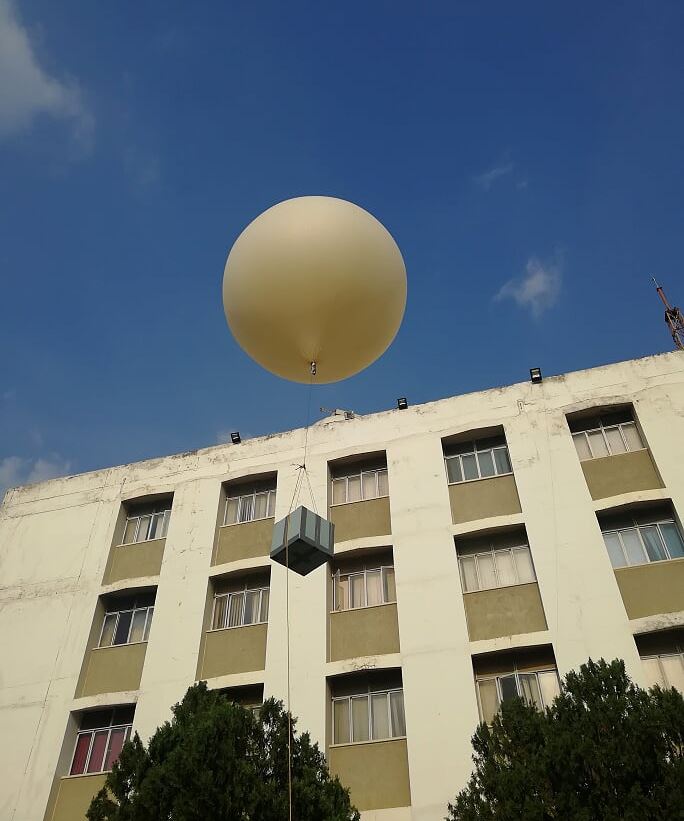}
 \hskip 0.1in
\includegraphics[width=.40\textwidth, height=0.45\textwidth]{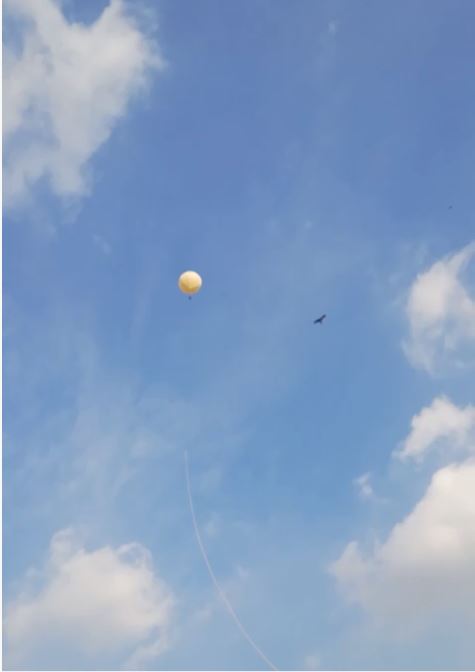}
\caption{{Tethered balloon flight on May 5th, 2019.}}
\label{tethered}
\end{figure}

Before the actual flight, a tethered flight, was performed on May 5th, 2019 from the Indian Institute of Astrophysics (IIA) campus, Bangalore to check the functioning of the CubeSat (Fig.~\ref{tethered}). CubeSat was placed in the Styrofoam box and wrapped in the bubble wrap for safety. The balloon was filled with $H_{2}$ gas. The payload was launched to an altitude of about 200 meters above the roof of the institute building. As expected, it took a while for the GPS to lock on to the satellites and, hence, we received very little data on latitude, longitude and altitude. However, all the other sensors were working properly. 

\begin{figure}[ht!]
\centering
 \includegraphics[width=.45\textwidth,height=0.45\textwidth]{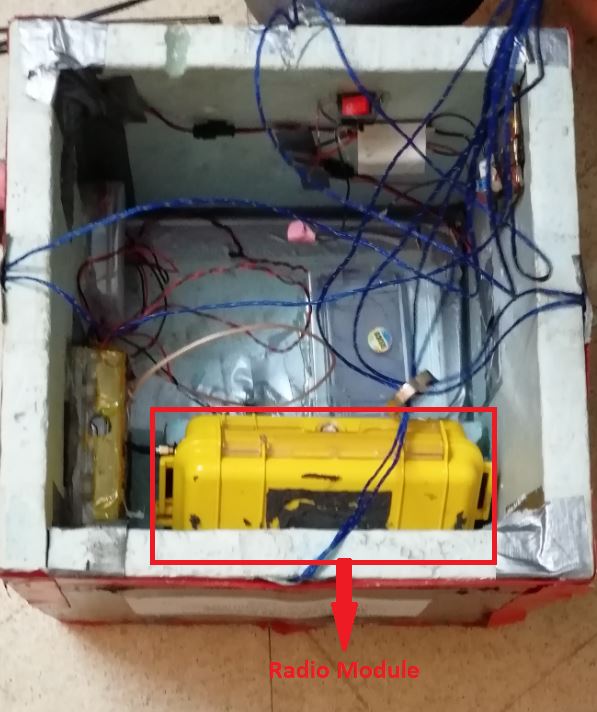}
 \hskip 0.1in
\includegraphics[width=.45\textwidth,height=0.45\textwidth]{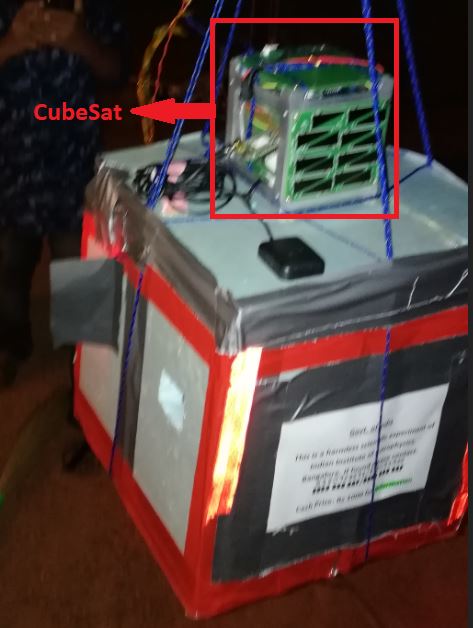} 
 \includegraphics[width=.45\textwidth, height=0.45\textwidth ]{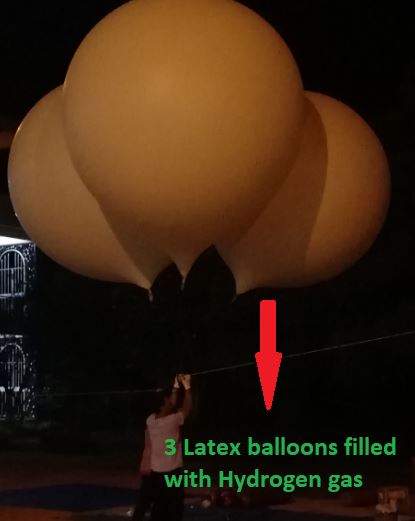}
 \hskip 0.1in
\includegraphics[width=.45\textwidth,height=0.45\textwidth]{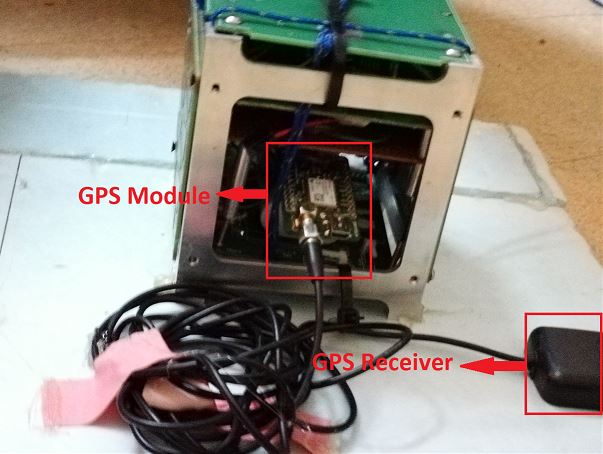}
\caption{Free-floating flight on March 8th, 2020. The CubeSat was placed on the top of the Styrofoam box (containing the instruments for another experiment and other electronic devices, like the radio module, etc.). Three balloons were inflated to carry the entire payload system for the experiment.}
\label{freefloat}
\end{figure}

The magnetometer data was used to check if the correct result was obtained. After applying Eq.~(\ref{finaleqnmagfield}) to the $X,Y,Z$ coordinates of the magnetic field of the Earth, we get the calibrated magnetic field as ($17.5956, -29.4674, 23.9948$) uT, respectively. The calibrated value of the total magnetic field was approximately 41.86 $\mu$T. The actual value of the total magnetic field in Bangalore as is 41.39 $\mu$T, which shows that our magnetometer was calibrated properly.\footnote{Verified with magnetic-declination.com/India/Bangalore/1132482.html}

The payload was finally launched to stratosphere from the CREST campus (13.133$^{\circ}$N, 77.815$^{\circ}$E) of the IIA, located in Hosakote, Bangalore. The entire balloon-payload system was launched at 3:09 am on March 8th, 2020. The total weight of the payload, including other components, such as a Geiger counter (for another experiment which details will be communicated separately), radio-module and other electronics, was about 4.9 kg. We used three latex balloons to lift the payload: two balloons of 1.2 kg-type and one 2-kg-type balloons were filled with hydrogen gas. The set up used for the free floating flight is shown in Fig.~\ref{freefloat}.

\section{Observations and Results}

\begin{figure}[ht!] 
\centering
\includegraphics[width=1\textwidth]{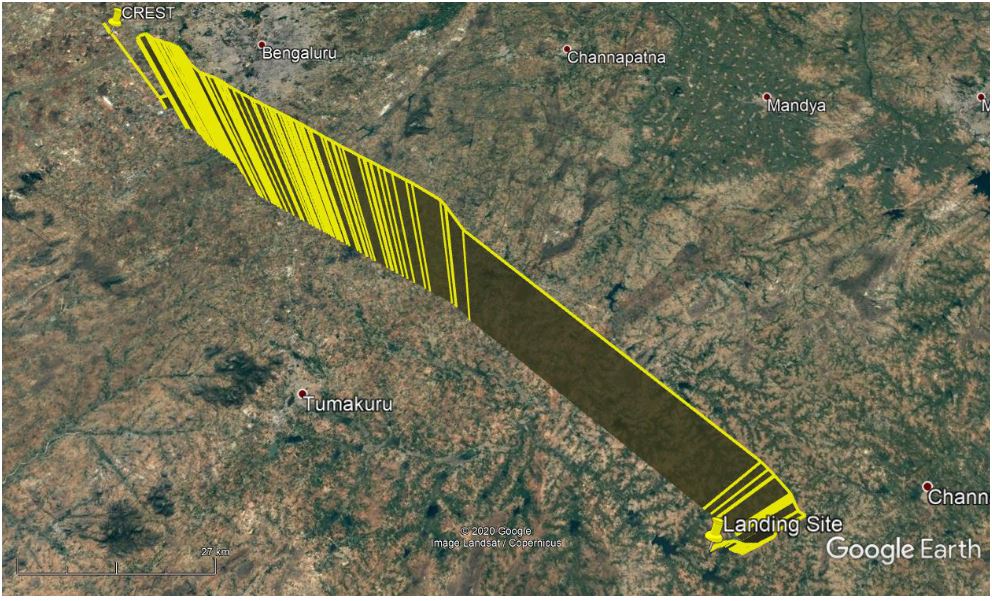} 
\includegraphics[width=1\textwidth]{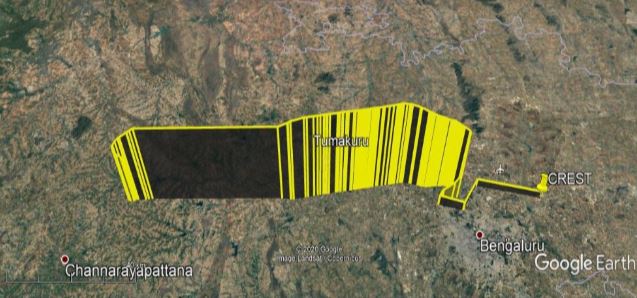}
\includegraphics[width=1\textwidth, height=0.25\textwidth]{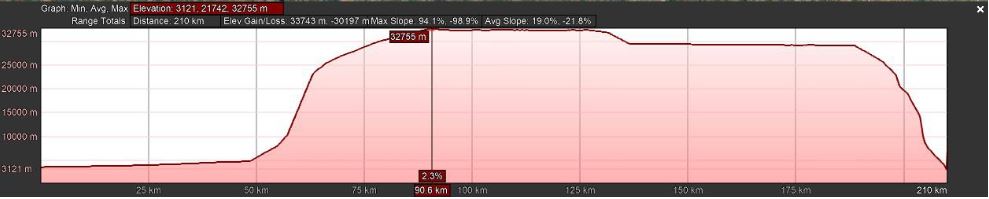}
\caption{The top two images represent the flight trajectory from the launch site (CREST) to the landing. The bottom image is a plot which represents the altitude of the payload as a function of the distance from CREST, the point of launch.}
\label{traj}
\end{figure}

During the flight, the data from the CubeSat were obtained only for about 1 hr. This translates to an altitude of about 11 km. The most probable reason for the short duration was because the battery was not insulated properly. We plan to rectify this in the subsequent flights. Here, we discuss the results of the obtained data. Though the CubeSat payload did not function for as long as we expected it to, we still got the data from the radio module. The payload landed at 13 14.4307N, 76 56.0865E in Tumkur district of Karnataka, at approximately 7:42 am. The maximum altitude reached was about 31 km, the lateral extent of the balloon path was about 96 kilometers, and the total duration of the flight was 4 hours and 24 minutes. The trajectory of the flight is shown in Fig.~\ref{traj}.

\begin{figure}[ht!]
\centering
 \includegraphics[width=.48\textwidth]{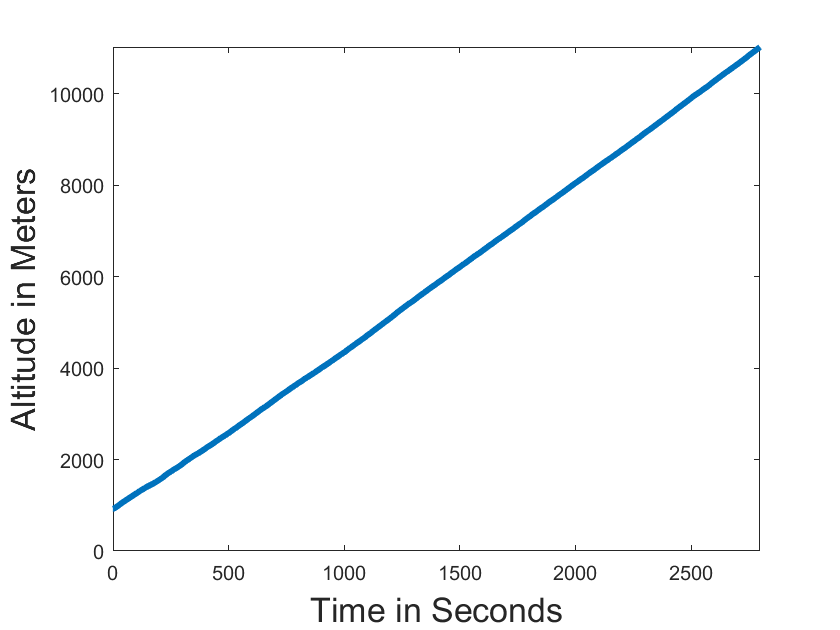}
 \hskip 0.1in
\includegraphics[width=.48\textwidth]{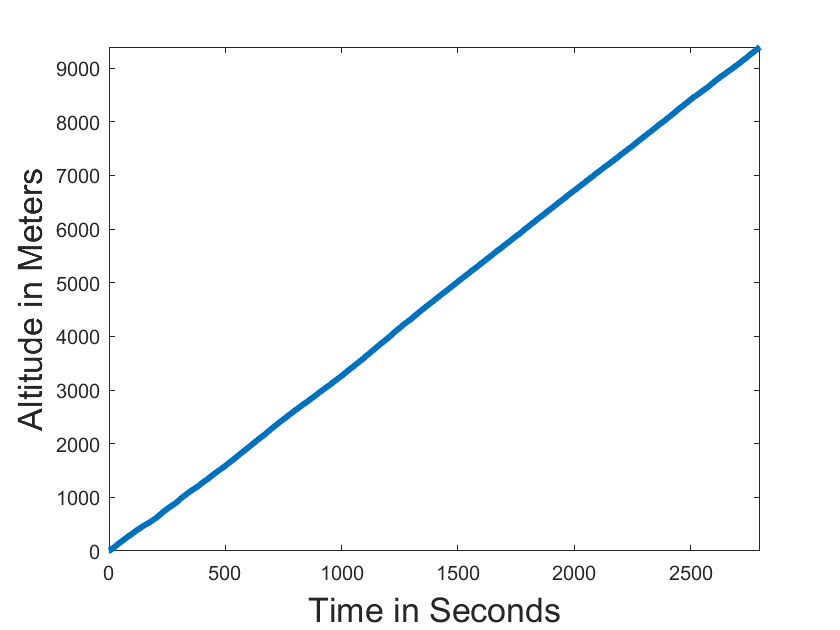} 
 \hskip 0.1in
\includegraphics[width=.48\textwidth]{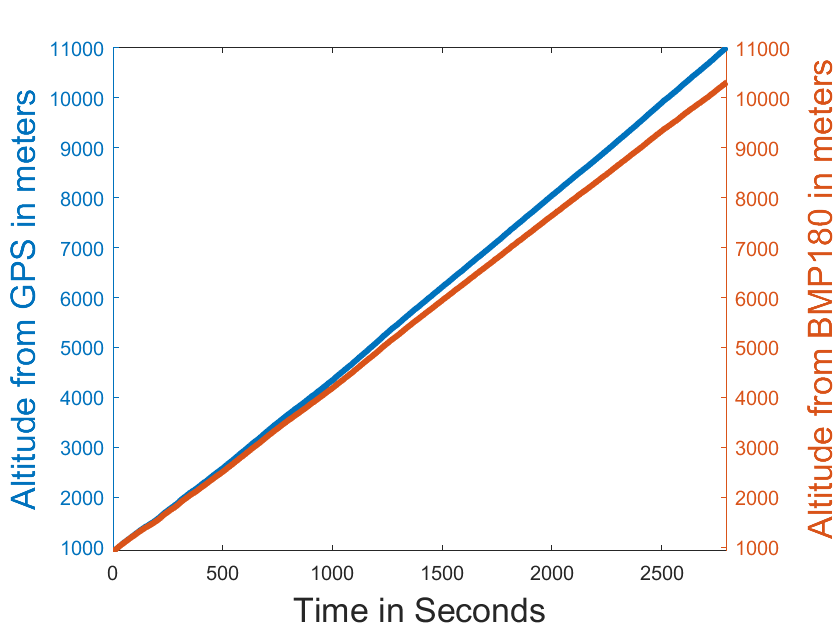} 
 \hskip 0.1in
\includegraphics[width=.48\textwidth]{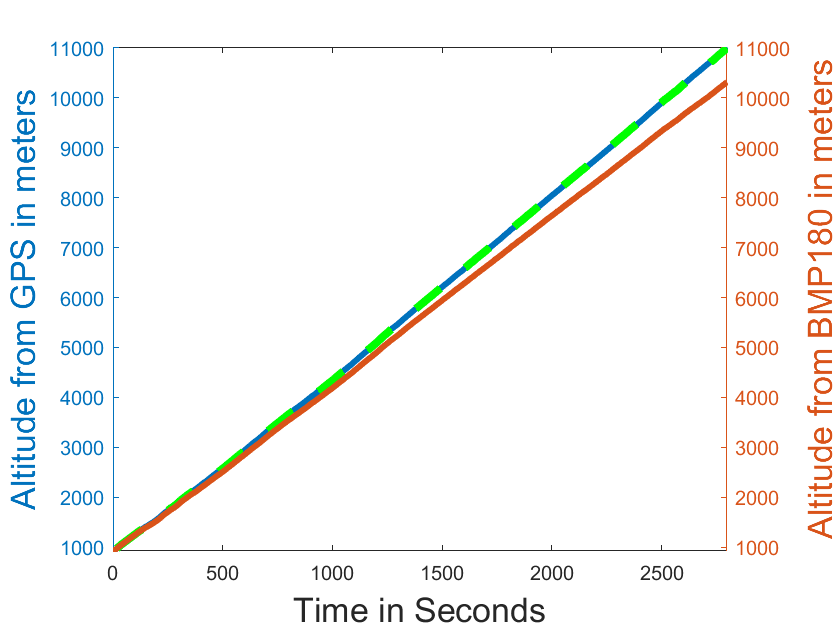} 
\caption{{\it Left}: The absolute altitude data estimated from the GPS. {\it Right}: The relative altitude estimated from the pressure reading obtained with the BMP180 pressure sensor. The third image shows the comparison of the altitude readings from the Pressure sensor and GPS. In the final image, the green dashed line represents the corrected altitude formula of Equation \ref{pressure}.}
\label{altvstime}
\end{figure}

We get the altitude data from two different sensors: the GPS and the BMP180 pressure sensor. The data given by the GPS is more accurate as we directly get the readings from the satellites. The data from the pressure sensor might not be as accurate as of the GPS, because we are indirectly estimating the data from the pressure reading using Eq.~\ref{pressure}. This estimate becomes less accurate as we go higher above the sea level. Another reason is that the temperature affects the pressure at any given point and, hence, affects the estimate of the altitude from the pressure. This is shown in Fig.~\ref{altvstime}. Note that the altitude vs time graph, estimated from the GPS, gives the absolute altitude above the sea level. Hence, the GPS data starts at about 950 meters -- the altitude of the CREST campus. \\ \newline
Further, in Fig.~\ref{altvstime}, we see that the overplotted image of the altitude vs. time obtained from the GPS does not match with the one obtained from BMP180. Since GPS is more accurate compared to BMP180, we modify Eq.~\ref{pressure} by fitting a curve to equation to the altitude obtained via GPS as:
\begin{equation}
A=66420 \left(1-\frac{p}{p_{o}}\right) ^ {0.1315}\,,
\label{pressure2}
\end{equation}
This modified equation (Eq.~\ref{pressure2}) is more accurate for low altitudes, however, its accuracy still needs to be further tested for higher altitudes. Also, the RMS of such a fit is just about 7.369, much better than Eq.~\ref{pressure}.\\

The altitude estimated from the pressure reading is the relative altitude from the CREST campus altitude. From these data, the average ascent speed of the flight was calculated to be about 3.68 m/s.  

\begin{figure}[ht!]
\centering
 \includegraphics[width=.48\textwidth]{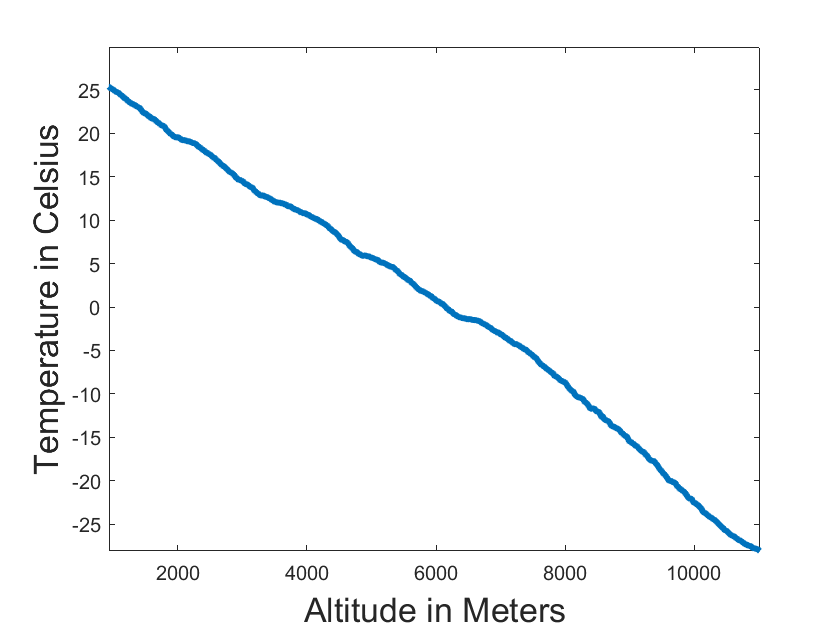}
 \hskip 0.1in
\includegraphics[width=.48\textwidth]{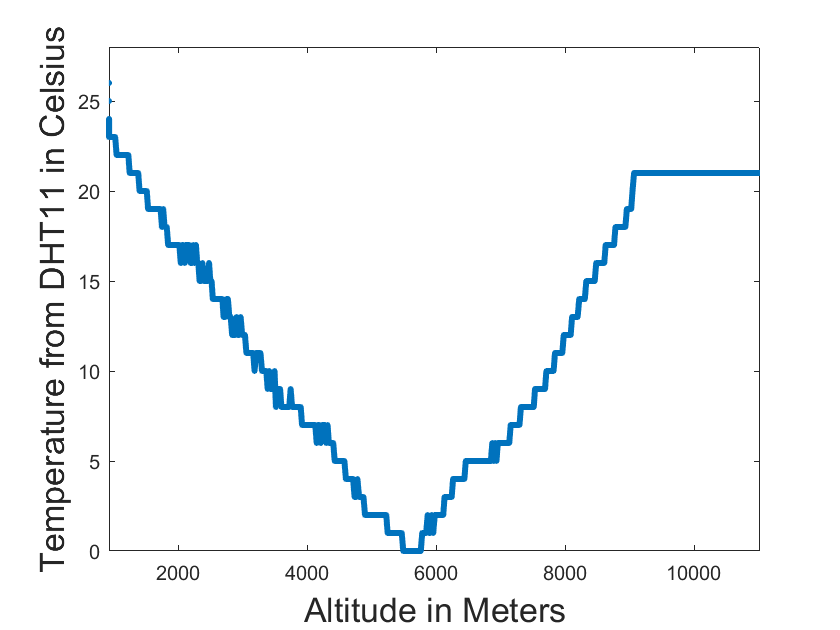} 
\caption{{\it Left}: Temperature measured using BMP180 sensor. {\it Right}: The temperature measured using DHT11 humidity sensor. We can see that it doesn't work properly below $0^{\circ}$C.}
\label{tempgraph}
\end{figure}

Next, the temperature was measured using the temperature sensor in the BMP180 pressure sensor and the temperature sensor in the DHT11 sensor Fig.~\ref{tempgraph}. The DHT11 temperature sensor did not work properly below $0^{\circ}$C as its operating range is $0^{\circ}$C to $50^{\circ}$C. Moreover, its precision and accuracy are lesser, and than the one on the BMP180 sensor. The temperature sensor in BMP180 can work up to $-40^{\circ}$C and can give results in decimal points. Hence, we prefer to use its data.

\begin{figure}[ht!]
\centering
 \includegraphics[width=.48\textwidth]{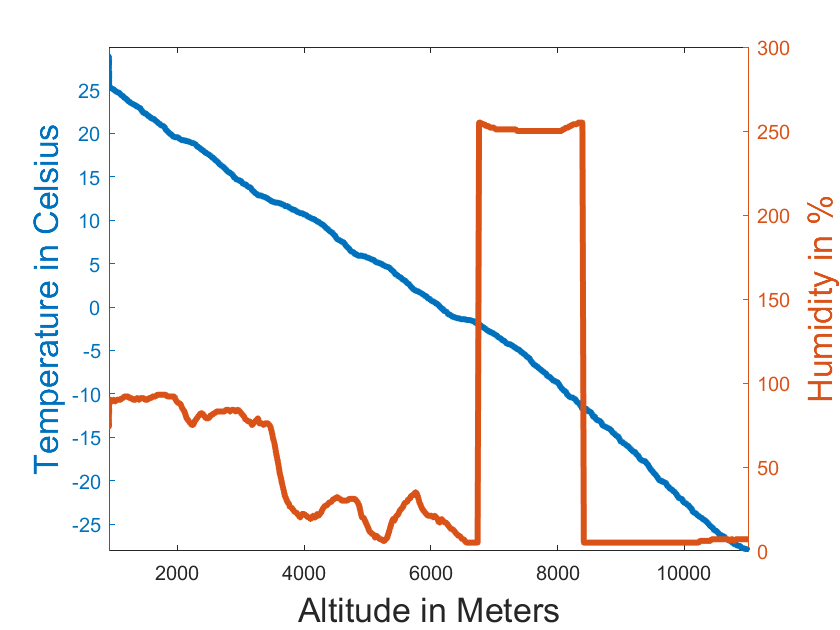}
 \hskip 0.1in
\includegraphics[width=.48\textwidth ]{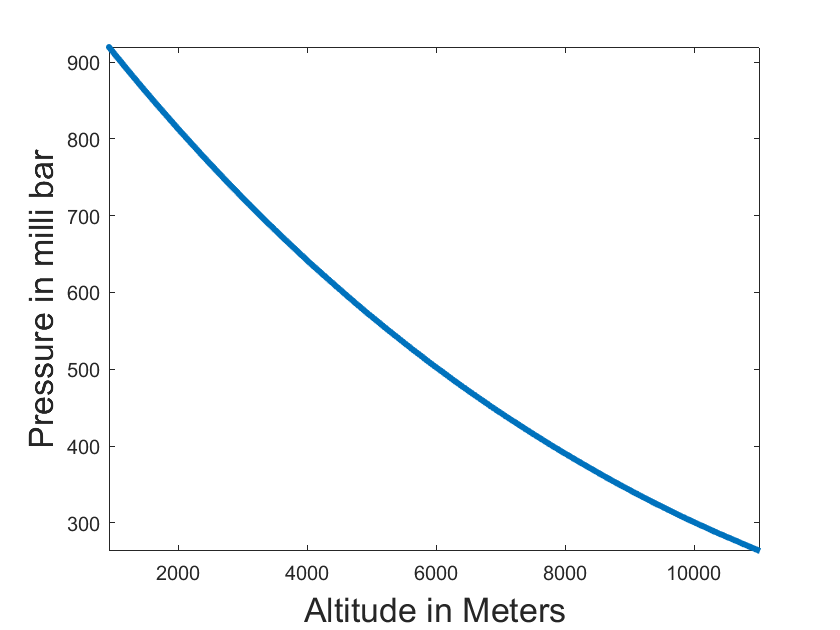} 
\caption{{\it Left}: Temperature and humidity data from DHT11. We can see a sudden increase in humidity at about $0^{\circ}$C. {\it Right}: Pressure (measured by BMP180) as a function of altitude. } 
\label{fig:humidity}
\end{figure}

Next, the humidity data measured in \% was plotted in Fig.~\ref{fig:humidity}. We see the sudden spike near the point where the temperature hits $0^{\circ}$C. We might have to investigate if this feature has to do anything with the freezing point of water.

The pressure seems to show a smooth drop as we increase in altitude. 

\begin{figure}[ht!]
 \includegraphics[width=.3\textwidth]{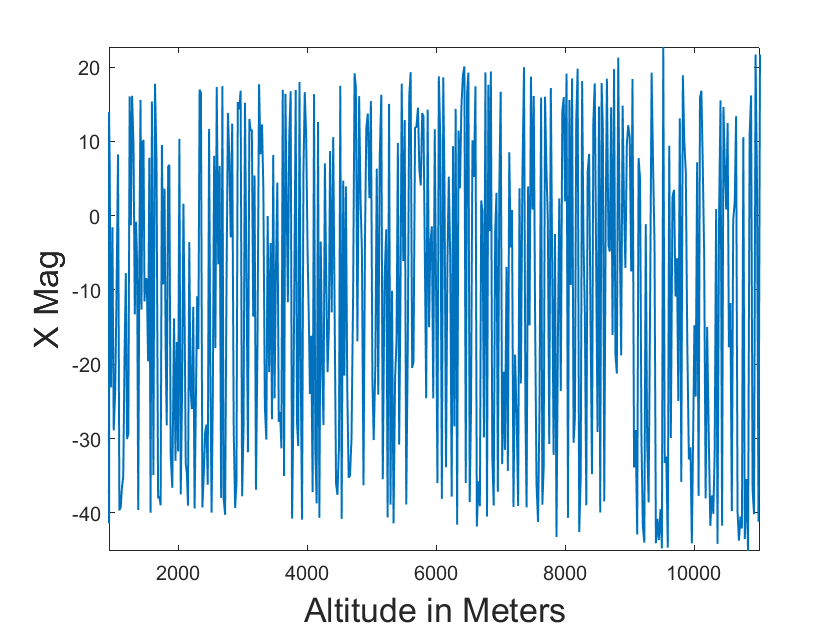}
\includegraphics[width=.3\textwidth]{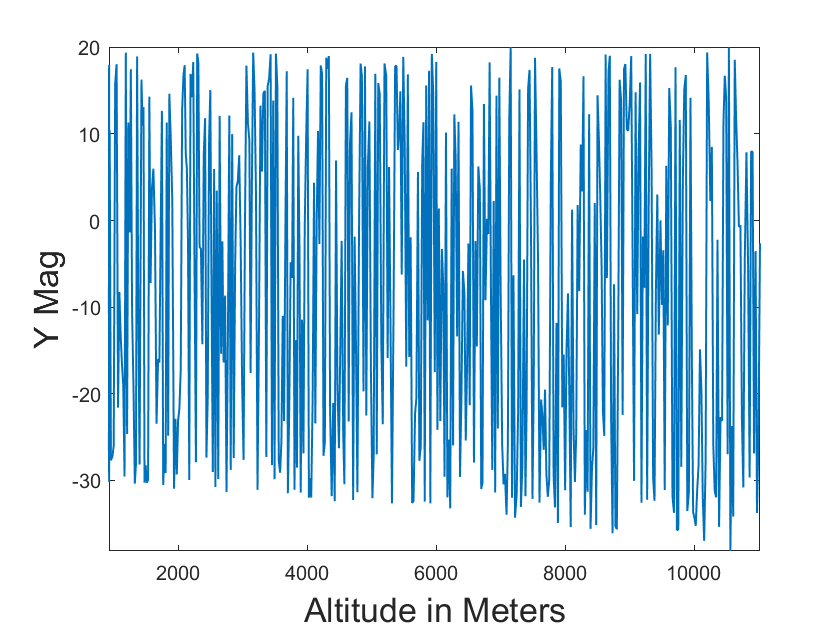} 
\includegraphics[width=.3\textwidth]{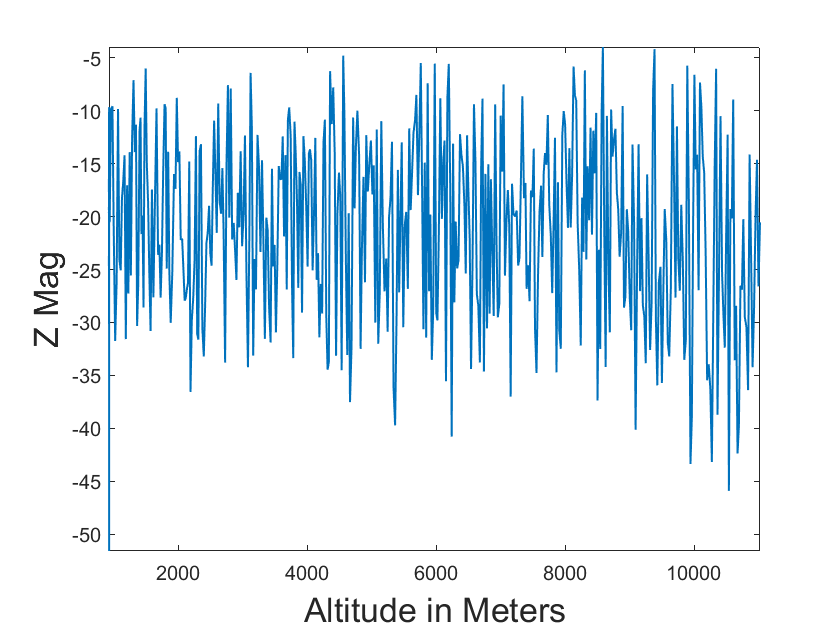} 
\vskip 0.05in
\includegraphics[width=.46\textwidth]{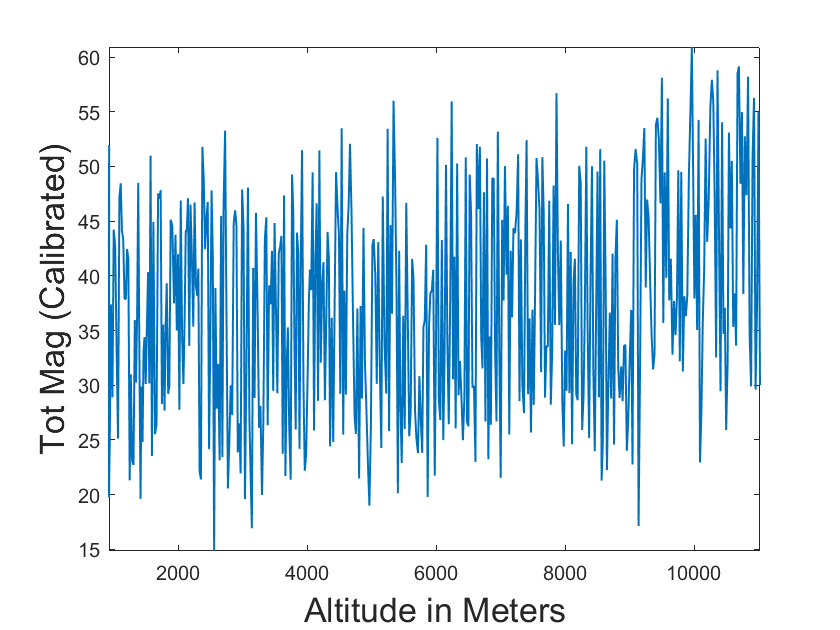}
\includegraphics[width=0.46\textwidth]{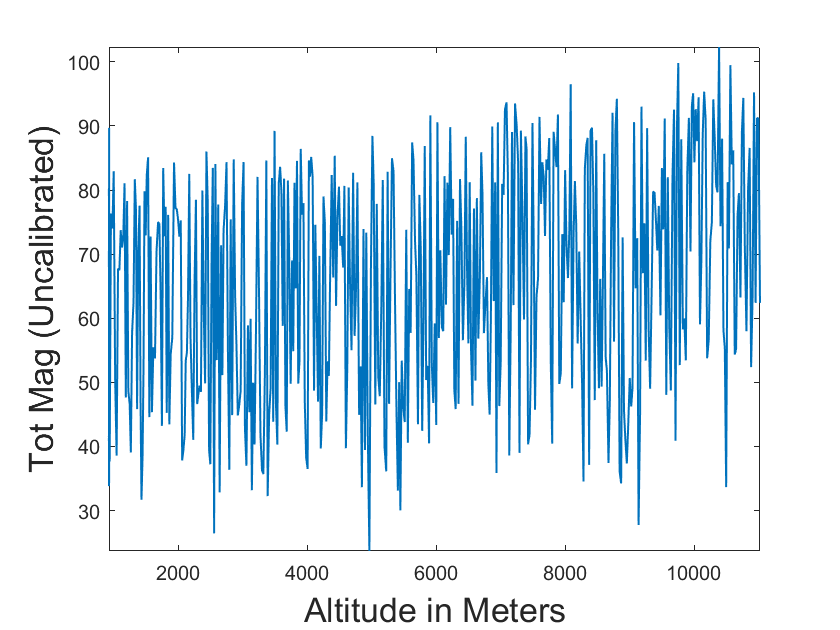}
\caption{{\it Top}: Plots of $X$, $Y$, and $Z$ components of the magnetic field. {\it Bottom}: Total magnitude of the uncalibrated magnetic field ({\it Left}) and calibrated magnetic field ({\it Right}). $X$ Mag, $Y$ Mag, $Z$ Mag and Tot Mag represents the $X$, $Y$, $Z$, and the total value of the magnetic field, respectively. Raw values of the magnetic field were changing rapidly with the altitude, and the total magnetic field may be showing some trend after 9000 meters altitude; however, further studies are required because of the insufficient data.}
\label{magfield}
\end{figure}

The $X$, $Y$ and $Z$ values of the magnetic fields are also plotted in Figure \ref{magfield}. The raw magnetic field obtained from the tethered flight after calibration seemed to give the correct magnetic field (Section 3). However, while plotting the data obtained from the free-floating launch, we had found that the total magnetic field keeps fluctuating with a mean of about 39 $\mu$T to 40 $\mu$T, between the 25 $\mu$T and 50 $\mu$T. To test it, we had plotted the uncalibrated magnetic field. To our surprise, the raw magnetic field obtained was varying very badly between 40 $\mu$T and 90 $\mu$T. If the raw values oscillate, then we cannot expect the actual magnetic field to be stable even if the calibration was performed correctly. Hence more studies need to be done on this. It is possible that the electronic components like the radio module might have induced some kind of magnetic field which is being sensed by the magnetometer. Further, there are several current-carrying wires, which in addition to producing its own magnetic field, also keeps rotating with the payload. This would further induce an additional magnetic field in the system. Because of these factors, we might have experienced an oscillating magnetic field. Thus it is essential that we perform more rigorous tests on the ground,  before we fly the entire system. The total magnetic field may be showing some trend after 9000 meters altitude. However, further studies are required because of the insufficient data.

\section{Summary and Conclusions}

Because we did not provide sufficient insulation, as soon as the temperature reached about $-27^{{\circ}}$C (which corresponds to an altitude of about 10 kilometers), the battery stopped functioning. The following can be concluded after studying the data from this flight:
\begin{itemize}
    \item Additional insulation has to be provided to the batteries so that they function even beyond 10 kilometers of altitude. If that still does not prove to be sufficient, we might have to develop a heating element with a feedback system. This might ensure that the payload is always at a constant temperature.
    \item It can be concluded that the magnetometer calibration has to be done in the presence of the entire payload system. In other words, the calibration has to be done in the presence of instruments, such as radio module and Geiger counter, which are going to be working throughout the duration of the flight. The fact that this was not done during the initial design, might have caused the total magnetic field value to fluctuate during the flight even though it did not happen during the tethered flight.
\end{itemize}
We plan to rectify these errors before the our next flight to further test the performance of CubeSat built from the off-the-shelf components.
  
\section{Acknowledgements}

Part of this research has been supported by the Department of Science and Technology (Government of India) under Grant IR/S2/PU-006/2012.


\begin{thebibliography}{}

\bibitem[Hibbits \textit{et al}.(2013)]{Ref-1} 
Hibbitts, C. A., Young, E., Kremic, T. and Landis, R. Science measurements and instruments for a planetary science stratospheric balloon platform. Aerospace Conference, IEEE, Big Sky, MT, pp. 1-9 (2013).

\bibitem[Nayak \textit{et al}.(2013)]{Ref1}
Nayak,~A., Sreejith, A.~G., Safonova,~M. and Murthy,~J. High-altitude ballooning programme at the Indian Institute of Astrophysics. Current Science, 104, 708, (2013).

\bibitem[Sreejith \textit{et al}.(2016)]{Ref2}
Sreejith A. G., Joice Mathew, Mayuresh Sarpotdar, Nirmal K., Ambily S., Ajin Prakash, Margarita Safonova, and Jayant Murthy. Balloon UV experiments for astronomical and atmospheric observations. Proc. SPIE 9908, Ground-based and Airborne Instrumentation for Astronomy VI, 99084E (9 August 2016).

\bibitem[Safonova \textit{et al}.(2017)]{Ref0} Safonova, M., Nayak, A., Sreejith, A.~G., Mathew, J., Sarpotdar, M., Ambily, S., Nirmal, K., Talnikar, S., Hadigal, S, Prakash. A. and Murthy, J. An Overview of High-Altitude Balloon Experiments at the Indian Institute of Astrophysics.  Astron. \& Astroph. Trans. (AApTr), 29(3): 397-426, (2016).

\bibitem[Mathew \textit{et al}.(2016)]{Ref3}
Mathew, J., Prakash, A., Sarpotdar, M., Sreejith, A. G., Safonova, M. and Murthy, J.  Ultraviolet cosmic imager to study bright UV sources. Paper 9905-146, Space Telescopes and Instrumentation 2016: Ultraviolet to Gamma Ray, part of SPIE Astronomical Telescopes + Instrumentation, (2016).

\bibitem[Sreejith \textit{et al}.(2016b)]{Ref5}
Sreejith, A. G., Mathew, J., Sarpotdar, M., Nirmal, K., Suresh, A., Prakash, A., Safonova, M., and Murthy, J. Measurement of limb radiance and Trace Gases in UV over Tropical region by Balloon-Borne Instruments – Flight Validation and Initial Results. Atmos. Meas. Tech. Discuss., https://doi.org/10.5194/amt-2016-98, (2016).

\bibitem[NASA(2017)]{C0}
NASA, Cubesat 101: Basic Concepts and Processes for First-Time CubeSat Developers. NASA and California Polytechnic State University, (2017).

\bibitem[Villela \textit{et al}.(2017)]{C1}
T. Villela, C. A. Costa, A. M. Brando, F. T. Bueno, and R. Leonardi. Towards the thousandth CubeSat: A statistical overview. Int. J. of Aerospace Engineering, vol. 5063145, no. 1, pp. 1–13, (Jan. 2017).

\bibitem[Techavijit \textit{et al}.(2016)]{C2} P. Techavijit, S. Chivapreecha, P. Sukchalerm and A. Plodpai. CubeSat image transmission in JPEG compression: An experiment on high altitude platform. 2016 8th International Conference on Knowledge and Smart Technology (KST), Chiangmai, pp). 164-168, (2016).

\bibitem[T{\o}mmer \textit{et al}.(2015)]{C3} M. T{\o}mmer, \textit{et al}. Testing of radio communication subsystems for the NUTS CubeSat on a meteorological balloon flight from Andoya in 2014. In Proceedings of the 22nd ESA Symposium on European Rocket and Balloon Programmes and Related Research, ESA Special Publication. ESA, (2015).

\bibitem[Kimm \textit{et al}.(2015)]{C3test} Kimm, H., Kang, J. S., Bruinga, B., and Ham, H. S. Real Time Data Communication Using High Altitude Balloon Based on Cubesat Payload. Journal of Advances in Computer Networks, 3(3), 186-190, (2015).

\bibitem[Mortensen \textit{et al}.(2010)]{C3_5} Mortensen, Hans Peter, \textit{et al}. NAVIS: Performance Evaluation of the AAUSAT3 Cubesat Using Stratospheric Balloon Flight. Proceedings of the 61th International Astronautical Congress: 3742–3749, (2010). 

\bibitem[Davis \textit{et al}.(2019)]{C4} J. Davis, \textit{et al}., Development of a High-Altitude Balloon CubeSat Platform for Small Satellite Education and Research, Small Satellite  Conference, (2019).

\bibitem[Kok \textit{et al}.(2012)]{I1} M. Kok, J. D. Hol, T. B. Sch\"{o}n, F. Gustafsson and H. Luinge. Calibration of a magnetometer in combination with inertial sensors. 15th International Conference on Information Fusion, Singapore, pp. 787-793, (2012).

\bibitem[Nirmal \textit{et al}.(2017)]{Nirmal} Nirmal, K., Sreejith, A. G., Mathew, J., Sarpotdar, M., Suresh Ambily, M. Safonova and J. Murthy. Pointing System for the Balloon-Borne Astronomical Payloads. Journal of Astronomical Telescopes, Instruments, and Systems (JATIS), 2(4), 047001, (2017). doi: 10.1117/1.JATIS.2.4.047001.

\end{thebibliography}
\end{document}